\documentclass[twocolumn,english,reprint,floatfix,amsmath,amssymb,aps,prl,superscriptaddress]{revtex4-1}
\usepackage{XCharter}
\usepackage[utf8]{inputenc}
\usepackage{geometry}
\usepackage{amsmath}
\usepackage{amssymb}
\usepackage{graphicx}

\makeatletter
\usepackage{dcolumn}\usepackage{bm}\usepackage{comment}\usepackage{url}\usepackage{XCharter}\usepackage{xspace}\usepackage{xr}\usepackage{cleveref}

\usepackage{babel}

\makeatother

\usepackage{babel}
\begin{document}
\title{Exact moments for a run and tumble particle with a finite tumble time
in a harmonic trap }
\author{Aoran Sun}
\email{Corresponding author. sunaoran16@mails.ucas.ac.cn}

\affiliation{Beijing National Laboratory for Condensed Matter Physics, Institute
of Physics, Chinese Academy of Sciences, Beijing 100190, China}
\affiliation{School of Physical Sciences, University of Chinese Academy of Sciences,
Beijing 100049, China}
\author{Fangfu Ye}
\email{Corresponding author. fye@iphy.ac.cn}

\affiliation{Beijing National Laboratory for Condensed Matter Physics, Institute
of Physics, Chinese Academy of Sciences, Beijing 100190, China}
\affiliation{School of Physical Sciences, University of Chinese Academy of Sciences,
Beijing 100049, China}
\affiliation{Wenzhou Institute, University of Chinese Academy of Sciences, Wenzhou,
Zhejiang 325001, China}
\author{Rudolf Podgornik}
\email{Deceased}

\affiliation{School of Physical Sciences, University of Chinese Academy of Sciences,
Beijing 100049, China}
\affiliation{Beijing National Laboratory for Condensed Matter Physics, Institute
of Physics, Chinese Academy of Sciences, Beijing 100190, China}
\affiliation{Wenzhou Institute, University of Chinese Academy of Sciences, Wenzhou,
Zhejiang 325001, China}
\affiliation{Kavli Institute for Theoretical Sciences, University of Chinese Academy
of Sciences, Beijing 100049, China}
\begin{abstract}
We study the problem of a \textsl{run and tumble particle} in a harmonic
trap, with a finite run and tumble time, by direct integration of
the equation of motion. An exact one dimensional (1D) steady state
distribution is obtained. Diagram laws and a programmable Volterra
difference equation are derived to calculate any order of moments
in any dimension, both for steady state as well as the time Laplace
transform for the intermediate states. We finally infer the complete
distribution from the moments, by considering a Gaussian quadrature
for the corresponding measure, and from the scaling law of higher
order moments. 
\end{abstract}
\maketitle

\section{Introduction}

Unlike Brownian particles that only exhibit net motion when passively
driven by an external force, active particles can move by making use
of energy provided by the environment, fueling their motion \citep{2013_HydrodynamicsofSoftActiveMatter,2016_ActiveParticlesinComplexandCrowdedEnvironments,2022_IrreversibilityandBiasedEnsemblesinActiveMatterInsightsfromStochasticThermodynamics}.
Many vivid examples of active particles can be found either in Nature,
such as molecular motors \citep{2006_ActiveGelsDynamicsofPatterningandSelfOrganization,2007_NonequilibriumMechanicsofActiveCytoskeletalNetworks},
cells \citep{2004_E.ColiinMotion,2008_OutofEquilibriumMicrorheologyinsideLivingCells},
granular materials \citep{2017_NoiseandDiffusionofaVibratedSelfPropelledGranularParticle},
active gels \citep{2010_TheMechanicsandStatisticsofActiveMatter,2019_SignaturesofMotorSusceptibilitytoForcesintheDynamicsofaTracerParticleinanActiveGel},
large (compared with cells) animals \citep{2004_AModeloftheFormationofFishSchoolsandMigrationsofFish,2005_HydrodynamicsandPhasesofFlocks,2014_FlockingataDistanceinActiveGranularMatter, Du_2021},
etc., or can be fabricated and exhibit robot-like qualities \citep{2014_GravitaxisofAsymmetricSelfPropelledColloidalParticles,2016_AcousticTrappingofActiveMatter,2016_ActiveParticlesinComplexandCrowdedEnvironments}.

Active particles have attracted substantial interest also theoretically,
due to nonequilibrium, non-Boltzmann statistics \citep{2014_GravitaxisofAsymmetricSelfPropelledColloidalParticles,2015_ActiveBrownianParticlesandRunandTumbleParticlesaComparativeStudy,2016_AcousticTrappingofActiveMatter},
even in the case of a single particle in free space. Run and tumble
particle (RTP) is one simple model that mimics the actual motion of
some bacteria, \textsl{e.g}., Escherichia coli \citep{2004_E.ColiinMotion,2022_WrappedUptheMotilityofPolarlyFlagellatedBacteria,Dobnikar}.
In this model, the active particle moves with constant velocity for
an exponentially distributed time (\textsl{the run state}), and then
randomly changes its velocity (\textsl{the tumble state}) to another,
randomly chosen velocity of the same magnitude (another run state).
Despite the apparent simplicity, such model already contains rich
features and can be non-trivial to analyze \citep{2015_ActiveBrownianParticlesandRunandTumbleParticlesaComparativeStudy,2020_RunandTumbleParticlesinTwoDimensionsMarginalPositionDistributions,2020_UniversalSurvivalProbabilityforadDimensionalRunandTumbleParticle}.
At a single particle level, time dependent distribution has been found
for the general case (in terms of its Fourier-Laplace transform) \citep{2013_AveragedRunandTumbleWalks}.
Other interesting quantities, such as first passage time \citep{2019_NoncrossingRunandTumbleParticlesonaLine,2020_UniversalSurvivalProbabilityforadDimensionalRunandTumbleParticle},
survival probability \citep{2020_UniversalSurvivalProbabilityforadDimensionalRunandTumbleParticle},
distribution of the time of maximum \citep{2019_GeneralisedArcsineLawsforRunandTumbleParticleinOneDimension},
have also been calculated. For many interacting particles, interesting
features including boundary clustering \citep{2016_ActiveParticlesinComplexandCrowdedEnvironments},
phase separation \citep{2015_MotilityInducedPhaseSeparation}, and
jamming \citep{2016_JammingandAttractionofInteractingRunandTumbleRandomWalkers},
have been observed.

An active particle, and more specifically, an RTP in an external potential,
is a natural and interesting generalization of this problem \citep{2015_RunandTumbleDynamicsofSelfPropelledParticlesinConfinement,2020_VelocityandDiffusionConstantofanActiveParticleinaOneDimensionalForceField,2024_ConfinedRunandTumbleParticleswithNonMarkovianTumblingStatistics}.
It may eventually reach a non-Boltzmann, nonequilibrium steady state
\citep{2022_ExactPositionDistributionofaHarmonicallyConfinedRunandTumbleParticleinTwoDimensions,2022_PositingtheProblemofStationaryDistributionsofActiveParticlesAsThirdOrderDifferentialEquation,2023_RunandTumbleOscillatorMomentAnalysisofStationaryDistributions,2024_ActiveOscillatorRecurrenceRelationApproach}.
For the special case of a harmonic potential, the exact steady state
distribution for RTP in one dimensional (1D) \citep{2015_PressureIsNotaStateFunctionforGenericActivefluids,2019_RunandTumbleParticleinOneDimensionalConfiningPotentialsSteadyStateRelaxationandFirstPassageProperties}
and 2D \citep{2022_PositingtheProblemofStationaryDistributionsofActiveParticlesAsThirdOrderDifferentialEquation,2023_RunandTumbleOscillatorMomentAnalysisofStationaryDistributions,2024_ActiveOscillatorRecurrenceRelationApproach}
have been found, as well as the moments of the steady state distribution
in the 3D case \citep{2022_PositingtheProblemofStationaryDistributionsofActiveParticlesAsThirdOrderDifferentialEquation,2023_RunandTumbleOscillatorMomentAnalysisofStationaryDistributions,2024_ActiveOscillatorRecurrenceRelationApproach}.

This model has been further generalized by including a random velocity
for run states \citep{2019_AFirstOrderDynamicalTransitionintheDisplacementDistributionofaDrivenRunandTumbleParticle},
a space-depending run rate \citep{2020_VelocityandDiffusionConstantofanActiveParticleinaOneDimensionalForceField,2020_RunandTumbleParticleinInhomogeneousMediainOneDimension},
a non-exponential time between tumbles \citep{2024_ConfinedRunandTumbleParticleswithNonMarkovianTumblingStatistics},
as well as a stochastical resetting to a starting point \citep{2018_RunandTumbleParticleunderResettingaRenewalApproach,2019_TelegraphicProcesseswithStochasticResetting}.

In general, the theoretical analysis of the RTP model, especially
in the presence of an external field, can be very challenging. Except
for the case when an exact solution is available, or a perturbation
analysis is applicable \citep{2020_TowardtheFullShortTimeStatisticsofanActiveBrownianParticleonthePlane,2021_DirectionReversingActiveBrownianParticleinaHarmonicPotential,2023_NonequilibriumSteadyStateofTrappedActiveParticles},
many problems still seem to be beyond what is currently feasible.
This seems to be at least in part due to the limitations of the theoretical
method. Currently the most common theoretical tool is the Fokker-Planck
equation. While it does successfully model the RTP well, it is often
difficult to solve, even numerically, especially in the presence of
a harmonic trap. 

We therefore naturally pose a question, whether alternative methods
exist and whether they are applicable to this problem. Guided by the
seminal work of Mark Kac \citep{1974_AStochasticModelRelatedtotheTelegraphersEquation},
we find that for the RTP model, it is possible to integrate the equation
of motion directly in order to obtain the moments, and from the moments,
it is possible to obtain a good estimation of the full distribution
density, or in some cases, even the exact distribution function. 

\begin{figure}[t]
\centering \includegraphics[width=0.475\textwidth]{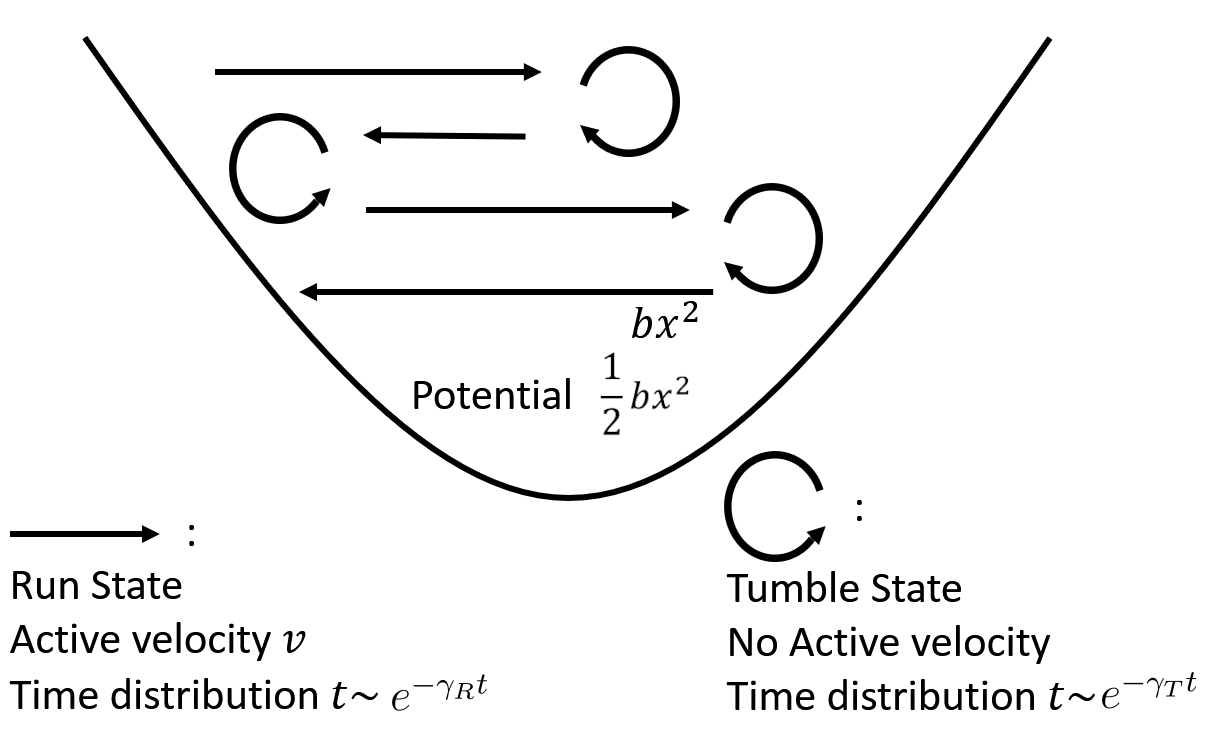} \caption{Schematic presentation of the 1D case of the run and tumble particle
with a finite tumble time in a harmonic trap. While the drawing pertains
to 1D for clarity, the method introduced here works for 2D and 3D
as well. The velocity of the particle has two components, the pull
$-bx$ of the harmonic potential $bx^{2}/2$, depending only on the
position, and the active velocity which switches randomly between
zero (\textsl{the tumble state}), and a vector of magnitude $v$ along
a randomly chosen direction (\textsl{the run state}). The time between
switches is exponentially distributed, with rate $\gamma_{R}$ and
$\gamma_{T}$, respectively for run and tumble state, and each choice
of the active velocity at the start of each run state is independent
of space, time, as well as previous choices.}
\label{fig:ad} 
\end{figure}

As these results for the standard RTP model are already obtained in
\citep{2022_PositingtheProblemofStationaryDistributionsofActiveParticlesAsThirdOrderDifferentialEquation,2023_RunandTumbleOscillatorMomentAnalysisofStationaryDistributions,2024_ActiveOscillatorRecurrenceRelationApproach}
\textsl{via} another method, we shall demonstrate our approach on
a variant of the standard RTP model, that is still beyond the reach
of other known methods, \textsl{i.e}., the RTP model with exponentially
distributed tumble time (see Fig. \ref{fig:ad} for the schematic
drawing of this model). Most of the recent theoretical approaches
assume that the tumble time is zero, \textsl{i.e}., the particle starts
another run immediately after one run, and thus always exhibits a
non-zero active velocity. In contrast, \textsl{Escherichia coli} is
observed to have a tumble state with small but nevertheless finite
time (typically 0.1s) between active runs (each typically 1s), during
which the cell mainly rotates and has nearly zero net motion \citep{2004_E.ColiinMotion}.
The finite tumble time is also important for the run-and-stop modeling
of R. sphaeroides \cite{2009_AMolecularBrakeNotaClutchStopstheRhodobacterSphaeroidesFlagellarMotor,2014_ModellingandAnalysisofBacterialTracksSuggestanActiveReorientationMechanisminRhodobacterSphaeroides}.
As for the BV2 cells, the typical tumble time, $11\min$, is much
longer than the typical run time, $1.8\min$ \cite{2023_RunandTumbleDynamicsandMechanotaxisDiscoveredinMicroglialMigration},
and therefore considering the finite tumble time is essential for
any reasonable modeling of these cells.

Despite the obvious importance of the finite tumble time, it is not
often discussed theoretically, partially because of the lack of an
adequate theoretical method. Indeed, in free space, when it is much
easier to consider, we see in \cite{2013_AveragedRunandTumbleWalks}
that a general tumble time is considered, and the time depending distribution
has been found in general (in terms of its Fourier-Laplace transform).
In the presence of an external potential the situation is altogether
different. The only theoretical result seems to be a 1D case with
a harmonic potential, where average times in run and tumble states
are equal \citep{2020_ExactStationaryStateofaRunandTumbleParticlewithThreeInternalStatesinaHarmonicTrap}.
This special case is essentially a direct product of two standard
1D RTPs projected back to 1D \citep{2022_ExactPositionDistributionofaHarmonicallyConfinedRunandTumbleParticleinTwoDimensions}.
We think it is highly possible that the restraint of equal average
run and tumble times, as well as the restriction to 1D, do not come
from the importance of this case, but are rather a reflection of the
limitation of currently available theoretical tools.

With our method, however, such a finite tumble time is rather straightforward
to implement. The approach is essentially the same as the standard
RTP model, with only a few differences stemming from the finite tumble
time. We derive the diagram laws to directly calculate the moments
for steady state, or the time Laplace transform for moments at any
time, with arbitrary run rate as well as tumble rate, in any dimension.
From the diagram laws we can find a Volterra difference equation \citep{2005_AnIntroductiontoDifferenceEquations}
which can be programmed as to recursively calculate the moments. Furthermore,
in 1D we can obtain an exact steady state distribution, and are thus
able to extend the result from \citep{2020_ExactStationaryStateofaRunandTumbleParticlewithThreeInternalStatesinaHarmonicTrap}
to the case when run time and tumble time are different.

The rest of the paper is organized as follows. In Section \ref{sec:Method-outline},
we define our model and describe our method. In Subsection \ref{subsec:Equation-of-motion}
we obtain one set of equations of motion for our model. We then briefly
describe the method to calculate the moments in Subsection \ref{subsec:Diagram-laws}.
The programmable Volterra difference equation is given in Subsection
\ref{subsec:Recursive-relation}, and in Subsection \ref{subsec:Remarks}
we consider some general results, such as the zero potential limit,
and the properties of the density at the boundary. We also briefly
consider the free space problem $b=0$ in Subsection \ref{subsec:FreeSpace},
and compare with the known results from \cite{2013_AveragedRunandTumbleWalks}.
In Section \ref{sec:1D}, we consider the special case of 1D, and
present an exact steady state distribution for arbitrary run time
and tumble time. In Section \ref{sec:Gauss=00003D00003D00003D000020rule}
we deduce the distribution from its moments. Finally, in Section \ref{sec:Conclusion}
we list the conclusions, where we summarize the results with a discussion
of the possible extensions based on the general methodology described
in the previous sections.

\section{Method outline }

\label{sec:Method-outline}

\subsection{Equation of motion}

\label{subsec:Equation-of-motion}

We consider a general motion in $D$ dimensional space, by focusing
on a single coordinate component, which contains the most important
information in the spherical symmetry. We model the RTP with finite
tumble time, during which the velocity of the particle has no active
component, but only a passive component from the external potential,
with the following equation:

\begin{equation}
\dot{x}\left(t\right)=-bx\left(t\right)+vF\left(J\left(t\right)\right)\label{eq:equation=00003D00003D00003D000020of=00003D00003D00003D000020motion}
\end{equation}
where $F\left(J\left(t\right)\right)$ or simply $F\left(t\right)$
stands for (the projection onto one coordinate axis of) the dimensionless
active velocity process, $x$ is the coordinate component of the position,
while $b$ is the strength of the harmonic trap and $v$ is the magnitude
of the active velocity, both considered as constants. $J$ is a two
state Markov process: 
\begin{equation}
R\underset{\gamma_{T}}{\overset{\gamma_{R}}{\rightleftharpoons}}T
\end{equation}
where we use $R$ and $T$ to represent the run and tumble state,
respectively. The rates $\gamma_{R}$ and $\gamma_{T}$, both considered
constants. As for $F$, we have $F\left(T\right)=0$ and $F\left(R\right)$
is randomly chosen according to the model of the active velocity.

In this way, both run and tumble have exponentially distributed time
with rate $\gamma_{R}$ and $\gamma_{T}$ respectively. For simplicity
we require $\mathbb{P}\left(J\left(0\right)=T\right)=\gamma_{R}/\left(\gamma_{R}+\gamma_{T}\right)$,
and as a consequence, $\mathbb{P}\left(J\left(t\right)=T\right)=\gamma_{R}/\left(\gamma_{R}+\gamma_{T}\right)$
for all $t\geqslant0$. 

The active velocities of the run state $F\left(R\right)$ are model
dependent. In general they could be sampled from any distribution
that has finite moments, but in this work we focus on the RTP, where
the (dimensionless) active velocity is randomly and uniformly chosen
from a unit sphere in $D$ dimension $S^{D-1}$. Since we are working
with only the $x$ coordinate component, we project the active velocity
onto the $x$ coordinate axis. The distribution for the square of
the projected active velocity $y=\left(F\left(R\right)\right)^{2}$
is: 
\begin{equation}
p\left(y\right)=\frac{\Gamma\left(\frac{D}{2}\right)\left(1-y\right)^{\frac{D-3}{2}}}{\sqrt{\pi y}\Gamma\left(\frac{D-1}{2}\right)}\label{eq:coordinatedist}
\end{equation}
This can be calculated as following: since the active velocity is
uniformly chosen from the unit sphere, the event $\mathbb{P}\left(\left(F\left(R\right)\right)^{2}<y\right)$
is proportional to the area of the spherical segment between hyperplanes
$z=\pm\sqrt{y}$ where $z$ is one coordinate. This area can be calculated
by integrating the area of a sphere in $D-1$ dimension with radius
$\sqrt{1-z^{2}}$ between $\pm\sqrt{y}$. Thus: 
\begin{equation}
\mathbb{P}\left(\left(F\left(R\right)\right)^{2}<y\right)\propto\int_{-\sqrt{y}}^{\sqrt{y}}\sqrt{\left(1-z^{2}\right)^{D-3}}dz
\end{equation}
Taking the derivative w.r.t $y$ and normalizing, we obtain Eq. \ref{eq:coordinatedist}.
Here, what we really need is the moments: 
\begin{equation}
M_{d}^{k}=\int y^{k}p\left(y\right)dy=\frac{\Gamma\left(1/2+k\right)\Gamma\left(d/2\right)}{\sqrt{\pi}\Gamma\left(d/2+k\right)}.
\end{equation}
For $d=1$ we have $M_{1}^{k}=1$ and thus the problem can be substantially
simplified. For $d=2$, what we need is basically: 
\begin{equation}
M_{2}^{k}=\frac{1}{2\pi}\int_{0}^{2\pi}\left(\cos^{2k}\theta\right)d\theta=\frac{\Gamma\left(1/2+k\right)}{\sqrt{\pi}\Gamma\left(1+k\right)}
\end{equation}
For $d=3$ the celebrated Archimedes' Hat-Box theorem \cite{2009_CRCEncyclopediaofMathematics}
indicates that $F$ is uniformly distributed over $\left[-1,1\right]$
and we thus have $M_{3}^{k}=1/\left(1+2k\right)$.

After a complete description of the model, we proceed to solving Eq.
\ref{eq:equation=00003D00003D00003D000020of=00003D00003D00003D000020motion}
as an ODE: 
\begin{align}
x\left(t\right) & =x\left(0\right)e^{-bt}+v\int_{0}^{t}F\left(t\right)e^{-b\left(t-s\right)}ds.\label{eq:solutionode}
\end{align}
A first observation from this solution is that, since $F\left(s\right)\leqslant1$,
$\left|x\left(t\right)\right|<v/b$ unless $\left|x\left(0\right)\right|>v/b$
and $t$ is small. Indeed, this can be proven by taking the absolute
value of Eq. \ref{eq:solutionode}: 
\begin{align}
\left|x\left(t\right)\right|\leqslant & \left|x\left(0\right)\right|e^{-bt}+v\int_{0}^{t}e^{-b\left(t-s\right)}ds\nonumber \\
< & \left|x\left(0\right)\right|e^{-bt}+\frac{v}{b}
\end{align}
With the solution Eq. $\ref{eq:solutionode}$, by interchanging the
order of the averaging and the integration, it is possible to calculate
the moments. As we are mostly interested in the long time behavior,
we therefore simplify the problem by assuming $x\left(0\right)=0$.
The moments of the probability distribution at time $t$ can then
be written generally as: 
\begin{widetext}
\begin{align}
\left\langle x\left(t\right)^{2l}\right\rangle = & v^{2l}\left(2l\right)!\int_{0\leqslant s_{1}\leqslant t_{1}\leqslant...\leqslant t_{l}\leqslant t}dt_{i}ds_{i}~e^{-b\left(2lt-\sum_{k}\left(t_{k}+s_{k}\right)\right)}\left\langle \prod_{k=1}^{l}~F\left(\left(t_{k}\right)\right)F\left(\left(s_{k}\right)\right)\right\rangle \label{momenteqsint}
\end{align}
where we have re-ordered the terms according to their time argument
and taking into account their symmetry. 
\end{widetext}

\subsection{Diagram laws}

\label{subsec:Diagram-laws}

The above integral is done in two steps. First, we shall calculate
the correlation function of the active velocities $\left\langle \prod_{k=1}^{l}~F\left(\left(t_{k}\right)\right)F\left(\left(s_{k}\right)\right)\right\rangle $,
then, with the insight of Kac, we shall perform the final integration
by using the Laplace transform.

To evaluate the correlation function, we use the law of total expectation
\cite{2013_StochasticProcesses}, or in other words, a case by case
discussion:  
\begin{equation}
\left\langle \cdot\right\rangle =\sum_{diag}\left\langle \cdot|diag\right\rangle \mathbb{P}\left(diag\right)
\end{equation}
Where $diag$ represent the cases under consideration, and $\mathbb{P}\left(diag\right)$
represents the probability that the case $diag$ happens. At any time,
$t_{i}$ and $s_{i}$, appearing in the correlation function, the
particle is either in the run state, or in the tumble state. We see
that if the particle is in the tumble state at any such time, then
$F\left(t_{i}\right)=0$ and naturally $\left\langle \cdot|diag\right\rangle =0$.
Therefore we may assume that the particle is in the run state for
all times appearing in the correlation function.

Then, for any two times appearing in the correlation function, either
the particle remains in the same run state the whole time, without
entering a tumble state, and thus the active velocity remains constant,
or the particle enters the tumble state at least once, and thus the
active velocities are uncorrelated. Each combination of these two
possibilities for all pairs of time forms one case to discuss, which
can be represented in the form of diagrams: we use a line of $2l$
vertices, each vertex representing a time appearing in the correlation
function. We connect two vertices if the particle remains in the same
run state at the two times they represents, or otherwise we leave
it blank. As an example, let us consider the following diagram, contributing
to the correlation function of order eight:  
\begin{equation}
\bullet_{t_{4}}-\bullet_{s_{4}}-\bullet_{t_{3}}-\bullet_{s_{3}}\quad\bullet_{t_{2}}-\bullet_{s_{2}}\quad\bullet_{t_{1}}-\bullet_{s_{1}}\label{eq:8th=00003D00003D00003D000020diag}
\end{equation}
which represents the case where $F$ remains in the same run state
from $s_{3}$ to $t_{4}$, from $s_{2}$ to $t_{2}$, and from $s_{1}$
to $t_{1}$, but enters the tumble state at least twice, first at
some time between $s_{2}$ and $t_{1}$, and again between $s_{3}$
and $t_{2}$. To calculate the contribution from one diagram, we need
the conditional expectation $\left\langle \cdot|diag\right\rangle $
and probability $\mathbb{P}\left(diag\right)$. The conditional expectation
is rather straightforward to calculate. Each diagram is broken by
the blanks into uncorrelated segments, thus the expectation can be
calculated by multiplying the expectations of each segment. Since
in each segment $F$ remains unchanged, the expectation simply equals
the moments. Therefore, each segment of length $2k$ contributes the
$2k$ moments $M_{d}^{k}$, and the total conditional expectation
is their product.

The probability can be calculated ``piece by piece'' according to
the Markov property, \textsl{i.e.}, by multiplying the following probabilities: 

1) $\mathbb{P}\left(J\left(s_{1}\right)=R\right)=\frac{\gamma_{T}}{\gamma_{R}+\gamma_{T}}$,
as $\left\langle \cdot|diag\right\rangle =0$ if $J\left(s_{1}\right)=T$. 

2) For any segment between $a$ and $b$ ($a\geqslant b$), the probability
that the particle remains in the same run state during that time equals
$e^{-\gamma_{R}\left(a-b\right)}$. This is easily obtained from the
exponentially distributed run time. 

3) For any blank between $a$ and $b$ ($a\geqslant b$), the probability
that the particle enters the tumble state at least once, but returns
to another run state at time $a$ equals:  
\begin{equation}
P\left(a,b\right)=\frac{\gamma_{T}}{\gamma_{R}+\gamma_{T}}+\frac{\gamma_{R}}{\gamma_{R}+\gamma_{T}}e^{-\left(\gamma_{R}+\gamma_{T}\right)\left(a-b\right)}-e^{-\gamma_{R}\left(a-b\right)},
\end{equation}
which can be obtained by solving the ODEs for $p_{R}\left(t\right)$
and $p_{T}\left(t\right)$, representing the probability to be in
the $R$ and the $T$ state at time $t$, respectively: 
\begin{equation}
\frac{d}{dt}p_{R}=-\gamma_{R}p_{R}+\gamma_{T}p_{T}
\end{equation}
\begin{equation}
\frac{d}{dt}p_{T}=-\gamma_{T}p_{T}+\gamma_{R}p_{R}
\end{equation}
with initial condition $p_{R}\left(b\right)=1,p_{T}\left(b\right)=0$.
We then calculating $p_{R}\left(a\right)$, to obtain the probability
to be in the $R$ state at time $a$. Finally we subtract $e^{-\gamma_{R}\left(a-b\right)}$
to exclude the case that the particle remains in the same $R$ state
without changing direction during time $a$ and $b$. 

As an example, the contribution from diagram Eq. \ref{eq:8th=00003D00003D00003D000020diag}
to the correlation function of order eight amounts to: 
\begin{equation}
\frac{\gamma_{T}e^{-\gamma_{R}\left(t_{2}-s_{2}+t_{1}-s_{1}\right)}P\left(s_{3},t_{2}\right)P\left(s_{2},t_{1}\right)e^{-\gamma_{R}\left(s_{4}-t_{3}\right)}M_{d}^{2}\left(M_{d}^{1}\right)^{2}}{\gamma_{R}+\gamma_{T}}
\end{equation}

What finally remains is to evaluate the required integrals in Eq.
\ref{momenteqsint}. All the integrands are products of exponential
functions, and are in fact convolutions. Noting the identity: 
\begin{align}
 & 2lt-\sum_{k=1}^{l}\left(t_{k}+s_{k}\right)\nonumber \\
= & 2l\left(t-t_{l}\right)+\sum_{k=1}^{l}\left(2k-1\right)\left(t_{k}-s_{k}\right)+\sum_{k=1}^{l-1}2k\left(s_{k+1}-t_{k}\right),\label{eq:integraltoconvolve}
\end{align}
one realizes that the Laplace transform of such an integral can be
easily calculated, and the result of the whole procedure is summarized
as the following diagram law: to calculate the Laplace transform of
the $2l$ moments:

1) draw $2l$ vertices on a line.

2) connect $2k$-th vertex with $2k-1$-th vertex for all $1\leqslant k\leqslant l$.

3) connect the $2k-1$-th vertex to the $2k-2$-th vertex for some
$1\leqslant k\leqslant l$.

4) calculate contribution from this diagram: If dot $m+1$ is connected
to dot $m$, factor 
\begin{equation}
\frac{m}{\gamma_{R}+mb+\xi}
\end{equation}
where $\xi$ is the Laplace transform variable. Otherwise factor 
\begin{align}
 & \frac{m\gamma_{T}\gamma_{R}}{\left(\gamma_{R}+\gamma_{T}+mb+\xi\right)\left(\gamma_{R}+mb+\xi\right)\left(mb+\xi\right)}.
\end{align}
In addition, every line passing through $2k$ dots gives a factor
$M_{d}^{k}$. Finally, there is also an overall factor 
\begin{equation}
\frac{2l\gamma_{T}v^{2l}}{\xi\left(2lb+\xi\right)\left(\gamma_{R}+\gamma_{T}\right)}.
\end{equation}
Multiplying everything together gives the contribution of this diagram.

5) sum over all possibilities in 3)  
\begin{widetext}
For the case of, \textsl{e.g.}, $l=2$ we need to consider two possible
diagrams: 
\begin{equation}
\bullet-\bullet-\bullet-\bullet,\qquad{\rm and}\qquad\bullet-\bullet\quad\bullet-\bullet
\end{equation}
and the corresponding moments can be obtained as 
\begin{align}
\widetilde{\left\langle x^{4}\right\rangle }\left(\xi\right)= & \frac{\gamma_{T}}{\gamma_{R}+\gamma_{T}}\frac{4v^{4}}{\xi\left(4b+\xi\right)}\frac{3}{\gamma_{R}+3b+\xi}\frac{2}{\gamma_{R}+2b+\xi}\frac{1}{\gamma_{R}+b+\xi}\frac{3}{d^{2}+2d}\nonumber \\
+ & \frac{\gamma_{T}}{\gamma_{R}+\gamma_{T}}\frac{4v^{4}}{\xi\left(4b+\xi\right)}\frac{3}{\gamma_{R}+3b+\xi}\frac{2\gamma_{T}\gamma_{R}}{\left(\gamma_{R}+\gamma_{T}+2b+\xi\right)\left(\gamma_{R}+2b+\xi\right)\left(2b+\xi\right)}\frac{1}{\gamma_{R}+b+\xi}\frac{1}{d^{2}}
\end{align}
\end{widetext}

In principle, all the moments are rational functions of $\xi$ and
thus the inverse Laplace transforms can be obtained. In reality, such
expressions would soon become unwieldy as one goes to higher order.
However, the limit $t\rightarrow\infty$, corresponding to the steady
state, can still be easily calculated by using the fact that: 
\begin{equation}
\lim_{t\rightarrow\infty}f\left(t\right)=\lim_{\xi\rightarrow0}\xi\widetilde{f}\left(\xi\right)
\end{equation}
The limit on the r.h.s. is trivial as can be seen from the above example.

\subsection{Volterra difference equation }

\label{subsec:Recursive-relation}

From the diagram law, we may derive a programmable Volterra difference
equation that avoids calculating all the $2^{l-1}$ diagrams for $2l$
moment and considerably simplifies the calculation.

It can be noticed that the part to the right (but not to the left)
of any blank in a diagram, is itself also a diagram, and a factor
of the whole diagram, so one can break any diagram from the leftmost
blank. Thus if one sets 
\begin{equation}
L^{l}\left(\xi\right)=\frac{\widetilde{\left\langle x^{2l}\right\rangle }\left(\xi\right)\xi\left(2lb+\xi\right)\left(\gamma_{R}+\gamma_{T}\right)}{2lv^{2l}\gamma_{T}}\label{equdef}
\end{equation}
one can obtain: 
\begin{widetext}
\begin{align}
L^{l}\left(\xi\right)= & \sum_{k=1}^{l-1}\left(\prod_{m=2k+1}^{2l-1}\frac{m}{\gamma_{R}+mb+\xi}\right)g_{k}L^{k}\left(\xi\right)M_{d}^{l-k}+\prod_{m=+1}^{2l-1}\frac{m}{\gamma_{R}+mb+\xi}M_{d}^{l}\label{eq:volterra=00003D00003D00003D000020deq}
\end{align}
where: 
\begin{equation}
g_{k}=\frac{2k\gamma_{T}\gamma_{R}}{\left(\gamma_{R}+\gamma_{T}+2kb+\xi\right)\left(\gamma_{R}+2kb+\xi\right)\left(2kb+\xi\right)}.
\end{equation}
\end{widetext}

The name \textsl{Volterra difference equation} comes from the fact
that it may be regarded as the discrete counterpart of the Volterra
integral equation. While difficult to solve in general, the Volterra
difference equation has the advantage that the r.h.s. requires only
$L^{k}$ up to $k=l-1$, and thus we can calculate the moment $\widetilde{\left\langle x^{2l}\right\rangle }\left(\xi\right)$
recursively. Starting with:

\begin{align}
L^{1}\left(\xi\right)= & \frac{1}{\gamma_{R}+b+\xi}M_{d}^{1}
\end{align}
we may then recursively calculate all $L^{k}$ up to $k=l$, and thus
recover the moment using Eq. $\ref{equdef}$.

\subsection{Remarks}

\label{subsec:Remarks}

So far we have been considering the distribution of one coordinate
components. Equivalently we are studying a RTP in a potential $\sim x^{2}/2$,
with active velocities in $d$ dimensions but the potential only in
one dimension. For $d>1$, one may wish to consider the distribution
of the vector $r$ instead of one of its coordinate components $x$.
Assuming spherical symmetry, we only need the distribution of $\left|r\right|$
or $\left|r\right|^{2}$. We could convert the moments for $x$ into
moments of $\left|r\right|^{2}$ as: $\left\langle \left(x\right)^{2l}\right\rangle =m_{d}^{l}\left\langle \left|r\right|^{2l}\right\rangle $,
where $m_{d}^{l}=\Gamma\left(1/2+l\right)\Gamma\left(d/2\right)/\left(\sqrt{\pi}\Gamma\left(d/2+l\right)\right)$
is the $2l$ moment of coordinate components for a random vector uniformly
chosen from a $d-1$ dimensional unit sphere. In our problem $m$
coincides with $M$ but in general they may be different. These are
already known from \citep{2022_PositingtheProblemofStationaryDistributionsofActiveParticlesAsThirdOrderDifferentialEquation,2023_RunandTumbleOscillatorMomentAnalysisofStationaryDistributions}
and we need not go further here.

One interesting feature of the RTP is that it may cluster near the
boundary. Here we examine such clustering on a single particle level,
which depends only on the rate of the run state $\gamma_{R}$. Numerically
we find that, for the steady state, as $l\rightarrow\infty$:

\begin{equation}
\left\langle \left(x\right)^{2l}\right\rangle \propto l^{-\gamma_{R}-\frac{d-1}{2}}\label{eq:scaling=00003D00003D00003D000020of=00003D00003D00003D000020moments}
\end{equation}

Notice that the last term in the Volterra difference equation $\prod_{m=+1}^{2l-1}\frac{m}{\gamma_{R}+mb}M_{d}^{l}$
has the same scaling law. The first implication is that, since the
moments are asymptotically decreasing, the distribution must be supported
in $\left[-1,1\right]$. Furthermore, by the same augment as in Ref.
\citep{2023_InteractingRunningandTumblingtheActiveDysonBrownianMotion},
as $x$ approaches the boundary, the density will approach: 
\begin{equation}
p\left(x\right)\propto\left(1-\left|x\right|\right)^{\gamma_{R}+\frac{d-1}{2}-1}.
\end{equation}
Therefore, we conclude that if $p\left(1\right)=0$ or $p\left(1\right)\rightarrow\infty$,
then $\gamma_{R}>\left(1-d\right)/2$ and $\gamma_{R}<\left(1-d\right)/2$,
respectively. For $d=2$ the critical value is $1/2$ and for $d=3$
it is $0$. Thus we may conclude that in 3D, the distribution of coordinate
component is not singular near the boundary. However, the distribution
of $r^{2}$ might still be singular.

\subsection{Free space}

\label{subsec:FreeSpace}

While we are considering RTP in a harmonic trap, it is simple to address
the free space limit by setting $b=0$, and obtaining the time-dependent
distribution in terms of Fourier-Laplace transform as in \citep{2013_AveragedRunandTumbleWalks}.
Start by defining:

\begin{equation}
K^{l}\left(\xi\right)=\frac{\widetilde{\left\langle x^{2l}\right\rangle }\left(\xi\right)\xi^{2}\left(\gamma_{R}+\gamma_{T}\right)}{\left(2l\right)!v^{2l}\gamma_{T}}
\end{equation}
\begin{equation}
f_{l}=\prod_{m=+1}^{2l-1}\frac{1}{\gamma_{R}+\xi}=\frac{1}{\left(\gamma_{R}+\xi\right)^{2l-1}}
\end{equation}
\begin{equation}
h=\frac{\gamma_{T}\gamma_{R}}{\left(\gamma_{R}+\gamma_{T}+\xi\right)\xi}.
\end{equation}
The Volterra difference equation for $K^{l}$ can be arranged into:
\begin{equation}
\frac{K^{l+1}}{f_{l+1}}=\sum_{k=0}^{l}\frac{1}{f_{k}}hK^{k}\left(\xi\right)M_{d}^{l+1-k}+M_{d}^{l+1}-\frac{hK^{0}\left(\xi\right)}{f_{0}}M_{d}^{l+1}.\label{eq:vdefree}
\end{equation}
The crucial simplification here is that $h$ does not depend on $k$,
unlike the case $b>0$, and thus this equation is of the convolution
type. Therefore it can be solved by means of the Z-transform \citep{2005_AnIntroductiontoDifferenceEquations}:

\begin{equation}
\hat{A}\left(z\right)=Z\left(A\left(n\right)\right)=\sum_{j=0}^{\infty}A\left(j\right)z^{-j}
\end{equation}
The explicit form of the Z-transform for $M_{d}^{l}$ is then obtained
as $\hat{M}_{d}$: 
\begin{equation}
\hat{M}_{d}\left(z\right)=Z\left(M_{d}^{l}\right)\left(z\right)=\frac{1}{d}~_{2}F_{1}\left(1,\frac{3}{2};1+\frac{d}{2};\frac{1}{z}\right)
\end{equation}
and it is possible to solve Eq. \ref{eq:vdefree} as

\begin{equation}
Z\left(\frac{K^{l}}{f_{l}}\right)\left(z\right)-\frac{K^{0}}{f_{0}}=\frac{~_{2}F_{1}\left(1,\frac{3}{2};1+\frac{d}{2};\frac{1}{z}\right)}{dz-h~_{2}F_{1}\left(1,\frac{3}{2};1+\frac{d}{2};\frac{1}{z}\right)}
\end{equation}
We then calculate the characteristic function by expanding and interchanging
the order of summation and expectation. Note that:

\begin{equation}
\sum_{l=1}^{\infty}\left(\frac{iqv}{\gamma_{R}+\xi}\right)^{2l}\frac{K^{l}}{f_{l}}=Z\left(\frac{K^{l}}{f_{l}}\right)\left(-\frac{\left(\gamma_{R}+\xi\right)^{2}}{q^{2}v^{2}}\right)-\frac{K^{0}}{f_{0}}
\end{equation}
It is therefore not necessary to carry out the inverse Z-transform,
and we can obtain directly:

\begin{equation}
\left\langle e^{iqx}\right\rangle =\frac{1}{\xi}-\frac{1}{\xi^{2}}\frac{p\left(\gamma_{R}+\xi\right)~_{2}F_{1}\left(1,\frac{3}{2};1+\frac{d}{2};-\left(\frac{qv}{\gamma_{R}+\xi}\right)^{2}\right)}{\frac{d\left(\gamma_{R}+\xi\right)^{2}}{q^{2}v^{2}}+h~_{2}F_{1}\left(1,\frac{3}{2};1+\frac{d}{2};-\left(\frac{qv}{\gamma_{R}+\xi}\right)^{2}\right)}.
\end{equation}

For the special case of $d=1,2,3$, it can be furthermore checked
that the above result reduces to \citep{2013_AveragedRunandTumbleWalks}.

\section{1D: exact Steady state distribution}

\label{sec:1D}

One specific case where the problem can be drastically simplified
is when $M_{d}^{l}=\left(M_{d}^{1}\right)^{l}$. The problem of 1D
run and tumble is one such case where $M_{1}^{l}=1$. In this case,
all non-zero diagrams of the same order have the same conditional
expectation regardless of the case $diag$, and Eq. \ref{eq:volterra=00003D00003D00003D000020deq}
then results in an explicit expression for any order of moments. Therefore,
by expanding $e^{iqx}$ and exchanging the order of summation and
expectation, we obtain the characteristic function for steady state
distribution in the form: 
\begin{equation}
\left\langle e^{iqx}\right\rangle =~_{1}F_{2}\left(\frac{\gamma_{T}}{2b};\frac{\gamma_{R}+\gamma_{T}}{2b},\frac{1}{2}+\frac{\gamma_{R}}{2b};-\frac{v^{4}q^{2}}{4b^{2}}\right)
\end{equation}

Where $_{1}F_{2}$ is the hypergeometric function. Noticing that $\left\langle e^{iqx}\right\rangle $
is the Fourier transform of the distribution function $p\left(x\right)$,
we can carry out the inverse Fourier transform. For simplicity we
rescale the time so that $b=1$, then rescale space so that $v=1$.
Under these conditions, the steady state distribution function becomes,
for $\left|x\right|<1$ (otherwise zero): 
\begin{widetext}
\begin{align}
p\left(x\right)= & \frac{\sqrt{\pi}\Gamma\left(\frac{\gamma_{R}+\gamma_{T}}{2}\right)\Gamma\left(\frac{1+\gamma_{R}}{2}\right)}{\Gamma\left(\frac{\gamma_{T}}{2}\right)\Gamma\left(\frac{\gamma_{R}}{2}\right)\cos\frac{\pi\gamma_{T}}{2}}\left\{ \frac{\left|x\right|^{\gamma_{T}-1}{}_{2}F_{1}\left(\frac{2-\gamma_{R}}{2},\frac{1-\gamma_{R}+\gamma_{T}}{2};\frac{1+\gamma_{T}}{2};x^{2}\right)}{\Gamma\left(\frac{1+\gamma_{R}-\gamma_{T}}{2}\right)\Gamma\left(\frac{1+\gamma_{T}}{2}\right)}-\frac{_{2}F_{1}\left(\frac{2-\gamma_{R}}{2},\frac{3-\gamma_{R}-\gamma_{T}}{2};\frac{3-\gamma_{T}}{2};x^{2}\right)}{\Gamma\left(\frac{\gamma_{R}+\gamma_{T}-1}{2}\right)\Gamma\left(\frac{3-\gamma_{T}}{2}\right)}\right\} \label{eq:exact=00003D00003D00003D000020solution}
\end{align}
\end{widetext}

\begin{figure*}
\centering \includegraphics[width=0.8\textwidth]{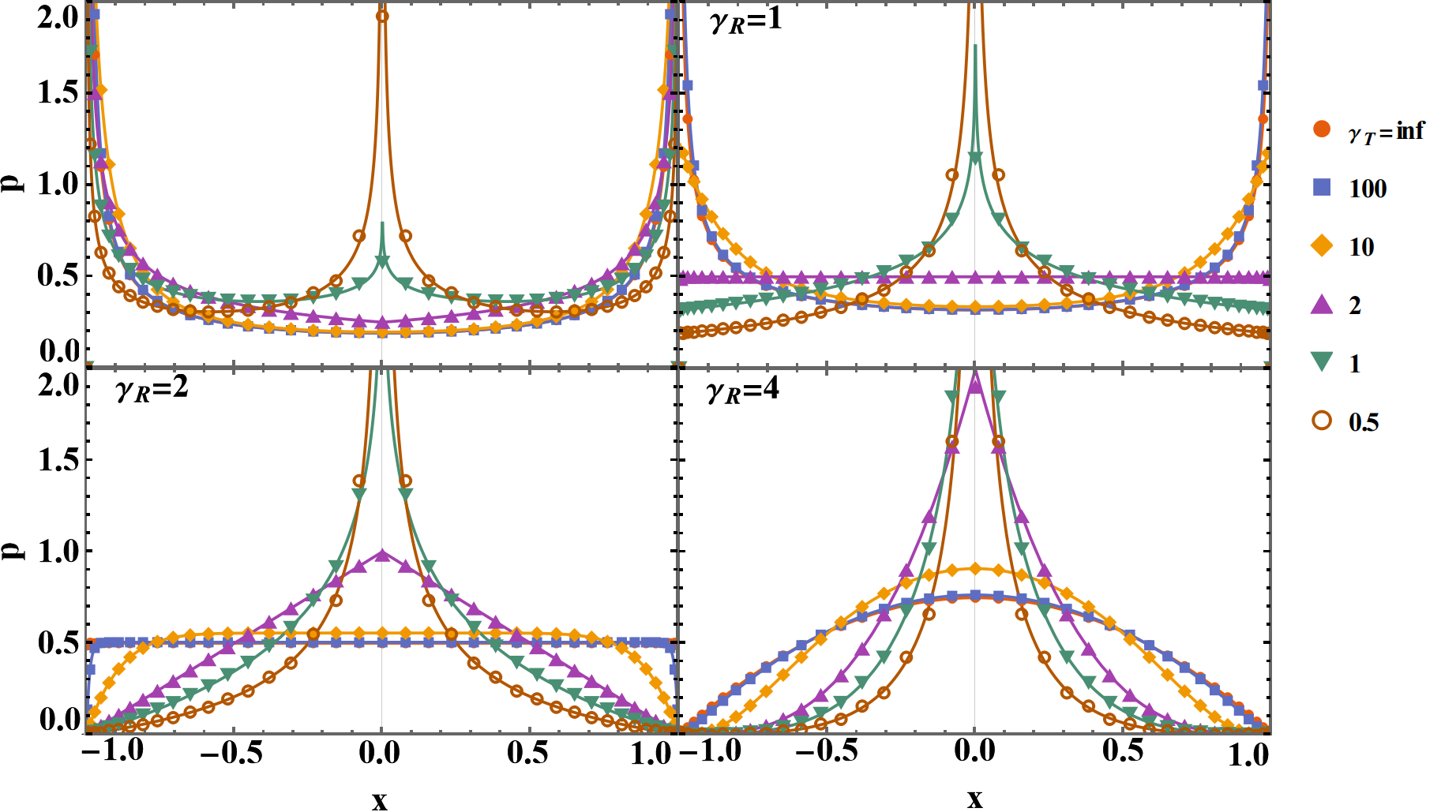} \caption{Examples of exact steady state distribution for a RTP in a harmonic
trap with finite tumble time in $1D$, according to Eq. \ref{eq:exact=00003D00003D00003D000020solution}
(solid line), and the estimated distribution with the method from
Section \ref{sec:Gauss=00003D00003D00003D000020rule} (dots). The
harmonic potential strength $b$ and the active velocity $v$ are
both rescaled to $1$. The run rates are $\gamma_{R}=1/2,1,2,4$ respectively.
In each graph the legend gives the value of the tumble rate $\gamma_{T}$.
The most apparent feature of the distribution is the peaks at the
center and the boundary. For the estimation, Chebyshev polynomials
with degrees up to $N=40$ are used. Even though $N$ is not very
large, the estimated distribution already agree with the exact distribution
well.}
\label{fig:1dex} 
\end{figure*}

First, in general, the hypergeometric function $_{2}F_{1}\left(\cdot,\cdot;\cdot;x\right)$
converges only for $\left|x\right|<1$. This is precisely the region
in which the particle will be bounded, as discussed in Subsection
\ref{subsec:Equation-of-motion}.

We then check our result by taking the limit $\gamma_{T}\rightarrow\infty$.
We work with the characteristic function since the distribution is
more complicated. We notice that with large $\gamma_{T}$, the first
two parameters of the characteristic function become identical, and
therefore the hypergeometric function will be reduced to

\begin{equation}
\left\langle e^{iqx}\right\rangle =~_{0}F_{1}\left(\frac{1+\gamma_{R}}{2};-\frac{q^{2}}{4}\right).
\end{equation}

We then perform the inverse Fourier transform and obtain:

\begin{equation}
p\left(x,\gamma_{T}\rightarrow\infty\right)=\frac{2\Gamma\left(\frac{3+\gamma_{R}}{2}\right)\left(1-x^{2}\right)^{\gamma_{R}/2-1}}{\sqrt{\pi}\left(1+\gamma_{R}\right)\Gamma\left(\frac{\gamma_{R}}{2}\right)}
\end{equation}

This agrees with the well known result in, \textsl{e.g}., \citep{2015_PressureIsNotaStateFunctionforGenericActivefluids,2019_RunandTumbleParticleinOneDimensionalConfiningPotentialsSteadyStateRelaxationandFirstPassageProperties},
except that $\gamma_{R}/2$ is the actual run rate as defined in some
references. The extra factor of $2$ does not appear in $D>1$.

Next we compare our results with \citep{2020_ExactStationaryStateofaRunandTumbleParticlewithThreeInternalStatesinaHarmonicTrap},
where it has been found that if $\gamma_{R}=\gamma_{T}>1$, the distribution
is continuous; whereas when $\gamma_{R}=\gamma_{T}<1$, the distribution
would exhibit three poles, located at the center and the two boundaries.
Our results here allow for the case $\gamma_{R}\neq\gamma_{T}$, and
it has been found that the singularities near the center and the boundaries
are controlled independently by $\gamma_{T}$ and $\gamma_{R}$, respectively,
as shown in Fig. \ref{fig:1dex}.

More specifically, $\gamma_{T}\leqslant1$ results in a singular peak
near the center, whereas for $\gamma_{T}>1,\gamma_{R}<\infty$, $p\left(0\right)$
is always finite, although it might be very large for large $\gamma_{R}$.
This can be shown from the expansion of $p$ near $0$, assuming $\gamma_{T}<2$:
\begin{eqnarray}
p\left(x\right) & \propto & \frac{\left|x\right|^{\gamma_{T}-1}}{\Gamma\left(\frac{1+\gamma_{R}-\gamma_{T}}{2}\right)\Gamma\left(\frac{1+\gamma_{T}}{2}\right)}-\nonumber \\
 &  & -\frac{1}{\Gamma\left(\frac{\gamma_{R}+\gamma_{T}-1}{2}\right)\Gamma\left(\frac{3-\gamma_{T}}{2}\right)}+O\left(x\right).
\end{eqnarray}
Thus the case $\gamma_{T}\neq1$ is obvious. Case $\gamma_{T}=1$
may be considered as the limit $\gamma_{T}\rightarrow1$: 
\begin{widetext}
\begin{equation}
p\left(x\right)\propto\left(_{2}F_{1}\left(\frac{2-\gamma_{R}}{2},\frac{2-\gamma_{R}}{2};1;x\right)\left(\ln\left|x\right|+\gamma_{E}+\frac{\Gamma'\left(\frac{\gamma_{R}}{2}\right)}{\Gamma\left(\frac{\gamma_{R}}{2}\right)}\right)+\left(_{2}F_{1}^{\left(0,1,0,0\right)}+{}_{2}F_{1}^{\left(0,0,1,0\right)}\right)\left(\frac{2-\gamma_{R}}{2},\frac{2-\gamma_{R}}{2},1,x\right)\right),
\end{equation}
\end{widetext}

It leads to a logarithmic singularity near the the center.

In addition, we find numerically that for $\gamma_{T}>2$, $p'\left(0\right)=0$,
whereas when $\gamma_{T}<2$, the derivative does not exist. In general,
it seems the $(k-1)$-th derivative at $0$ exists iff $\gamma_{T}>k$,
while the odd order derivative is zero due to symmetry.

Similarly, $\gamma_{R}<1$ results in singular peaks near the boundaries,
whereas for $\gamma_{R}>1$, $p\left(1\right)=0$. This can be proved
by the scaling law of moments Eq. \ref{eq:scaling=00003D00003D00003D000020of=00003D00003D00003D000020moments}.
If $\gamma_{R}=1$ the behavior will depend on $\gamma_{T}$: for
$\gamma_{T}<2$, $p\left(1\right)$ is finite, whereas for $\gamma_{T}>2$,
it diverges. For $\gamma_{R}=1,\gamma_{T}=2$, one has a uniform distribution
$p\left(x\right)=1/2$. In addition, the derivatives at boundaries
exhibit a similar behavior to the derivatives at center: the $k-1$
derivative at $1$ exists iff $\gamma_{R}>k$. Furthermore, when the
derivative exists, it is always zero.

Unfortunately, such tricks are almost exclusive applicable to 1D,
since from the requirement $M_{d}^{l}=\left(M_{d}^{1}\right)^{l}$,
one can calculate its characteristic function and after the inverse
Laplace transform one can show such distribution is the Bernoulli
distribution. Therefore, the method explained here is applicable to
1D or to a simple direct product of 1D (like in \citep{2022_ExactPositionDistributionofaHarmonicallyConfinedRunandTumbleParticleinTwoDimensions}).

Finally, we note that although avoiding the Fokker-Planck equation
is one major motivation of this work, it is nevertheless interesting
to check if our result agrees with the Fokker-Planck equation. Unfortunately,
here we only obtain the total distribution of particles in all states,
whereas for the Fokker-Planck equation we need the partial distribution
of particles in each state, running in each direction. The information
here is still insufficient, although with this additional information
about the total distribution it is possible to solve the Fokker-Planck
equation at least for some special values of the parameters. We shall
defer more discussion to a future work, where by extending the present
methodology we are able to work out all the necessary partial distributions,
and verify directly that they indeed solve the Fokker-Planck equation.

\section{Steady state distribution: Gaussian quadrature}

\label{sec:Gauss=00003D00003D00003D000020rule}

Without the simplification in Section \ref{sec:1D}, the exact solutions
are difficult to find. Yet we may still approximate the distribution
from its finitely-many moments that can be calculated at least numerically.
Extracting information about distribution from its moments is a century-old
problem in mathematics referred to as \textsl{the moment problem}
\citep{2020_TheClassicalMomentProblemandSomeRelatedQuestionsinAnalysis}.
While theories about the existence of the distribution and some estimations
in terms of inequalities can be found, it seems that an adequate approach
in practice is still lacking. In \citep{2024_ActiveOscillatorRecurrenceRelationApproach}
the practical problem was solved by expanding the distribution in
terms of the Legendre polynomials. However, such approach only works
well when the distribution is smooth, whereas in our case, the distribution
may be singular, resulting in a slow convergence.

Thus we approach this problem via another route. Briefly, knowing
the distribution is essentially the same as knowing the expectation
of an arbitrary function. Since our distribution has closed support,
any smooth function can be approximated well by the Chebyshev polynomials.
The expectation of the Chebyshev polynomials, since polynomials are
summations of monomials or powers, can be calculated from the moments.
Thus we may move backwards, from moments we know the expectation of
the Chebyshev polynomials, then the expectation of any smooth function,
and finally the distribution itself.

We rescale the time and length such that $b=v=1$. Then $x$ will
lie between $-1$ and $1$. An arbitrary smooth function $f$ supported
within $\left[-1,1\right]$ can be well approximated by summation
of finitely many Chebyshev polynomials. This approximation is almost
ideal due to its exponential convergence rate, explicit grids, minimal
amplitude and thus low uniform error (compared with other orthonomal
polynomials) \citep{2013_ChebyshevandFourierSpectralMethodsSecondRevisedEdition}.

The Chebyshev approximation can be explicitly constructed by the values
of $f$ on the Gauss-Lobatto grids: define $x_{i}=\cos\frac{\pi i}{N}$;
$p_{i}=2$ if $i=0$ or $i=N$, otherwise $p_{i}=1$; $J_{ij}=\frac{2}{p_{i}p_{j}N}\cos\frac{\pi ij}{N}$,
then the Chebyshev approximation can be written as 
\begin{equation}
f\left(x\right)\thickapprox\sum_{i,j=0}^{N}J_{ij}f\left(x_{j}\right)T_{i}\left(x\right)
\end{equation}
where $T_{n}\left(\cos\theta\right)=\cos n\theta$ is the $n$-th
Chebyshev polynomial (of the first kind). Taking the average on both
sides, we have:

\begin{equation}
\left\langle f\right\rangle \thickapprox\sum_{i,j=0}^{N}J_{ij}\left\langle T_{i}\left(x\right)\right\rangle f\left(x_{j}\right)
\end{equation}
This gives us the Gaussian quadrature to evaluate the expectation
of any smooth function, with the abscissas being $x_{j}$ and the
weights being $w_{j}=\sum_{i=0}^{N}J_{ij}\left\langle T_{i}\left(x\right)\right\rangle $.
In our problem, the expectation of Chebyshev polynomials can be evaluated
exactly from the moments by writing the Chebyshev polynomials as the
summation of the monomials or powers, then exchange the order of summation
and expectation, and thus the error comes only from the Chebyshev
approximation, which decays exponentially and uniformly over the whole
region.

On the other hand, suppose we could continuously extend $x_{j}$ such
that $j$ is allowed to be a real number. In that case we have: 
\begin{equation}
\left\langle f\right\rangle =\int f\left(x\left(j\right)\right)p\left(x\left(j\right)\right)\frac{dx}{dj}dj,
\end{equation}
assuming that $p$ is smooth near $x_{j}$. By approximating the integral
with finite sum, we have:
\begin{equation}
\left\langle f\right\rangle \thickapprox\sum_{j=0}^{N}f\left(x_{j}\right)p\left(x\left(j\right)\right)\frac{dx}{dj}
\end{equation}
Therefore, comparing with the Chebyshev approximation, we arrive at:
\begin{equation}
\sum_{j=0}^{N}f\left(x_{j}\right)p\left(x\left(j\right)\right)\frac{dx}{dj}\thickapprox\sum_{i,j=0}^{N}J_{ij}\left\langle T_{i}\left(x\right)\right\rangle f\left(x_{j}\right)
\end{equation}
As this approximation holds for arbitrary smooth $f$, we may consider
a situation where $f\left(x_{j}\right)=\delta_{jk}$, therefore:

\begin{equation}
p\left(x_{k}\right)\frac{dx_{k}}{dk}\thickapprox\sum_{i=0}^{N}J_{ik}\left\langle T_{i}\left(x\right)\right\rangle 
\end{equation}
Since we know $\left|\partial_{i}x_{i}\right|=\frac{\pi}{N}\left|\sin\frac{\pi i}{N}\right|$,
we obtain an approximate distribution at discrete points:

\begin{equation}
p\left(\cos\frac{\pi k}{N}\right)\sim\sum_{i=0}^{N}\frac{J_{ik}\left\langle T_{i}\left(x\right)\right\rangle }{\frac{\pi}{N}\left|\sin\frac{\pi k}{N}\right|}\label{eq:ChebyApprox}
\end{equation}
Eq. \ref{eq:ChebyApprox} is the main result of our approach to the
moment probem. It provides an explicit estimation of the density,
using only the expectation of the Chebyshev polynomials, which can
be calculated from moments. To validate Eq. \ref{eq:ChebyApprox},
we first compare the estimated distribution with the exact solution
in 1D. As shown in Fig. \ref{fig:1dex}, while we use a relatively
small $N=40$ for clarity, and the distributions themselves are complicated,
some with multiple singularities, the method works well to capture
the distribution even close to the singularity. While it is still
not able to recreate the singularity well, it does offer a hint of
possibility of a singular peak. 

\begin{figure*}
\centering \includegraphics[width=0.8\textwidth]{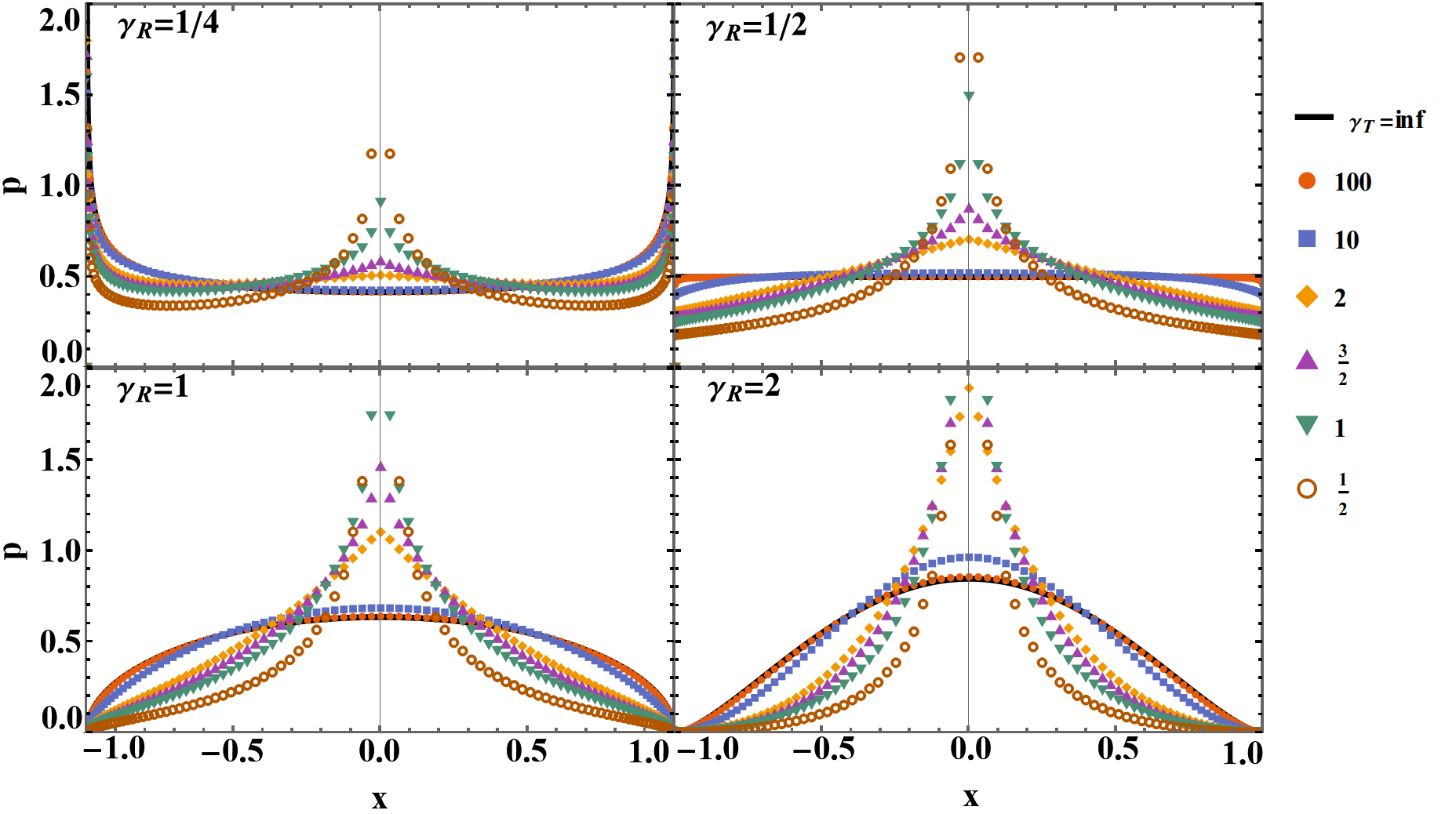}\caption{Approximated distribution from the Gaussian quadrature for a RTP in
the harmonic trap with finite tumble time in $2D$ with $N=100$ moments.
The harmonic potential $b$ and the active velocity $v$ are rescaled
to $1$. The run rates are $\gamma_{R}=1/4,1/2,1,2$ respectively.
In each graph the legends gives the value of $\gamma_{T}$. The solid
lines represent the exact distribution for $\gamma_{T}\rightarrow\infty$ }
\label{fig:2d} 
\end{figure*}

\begin{figure*}
\centering \includegraphics[width=0.8\textwidth]{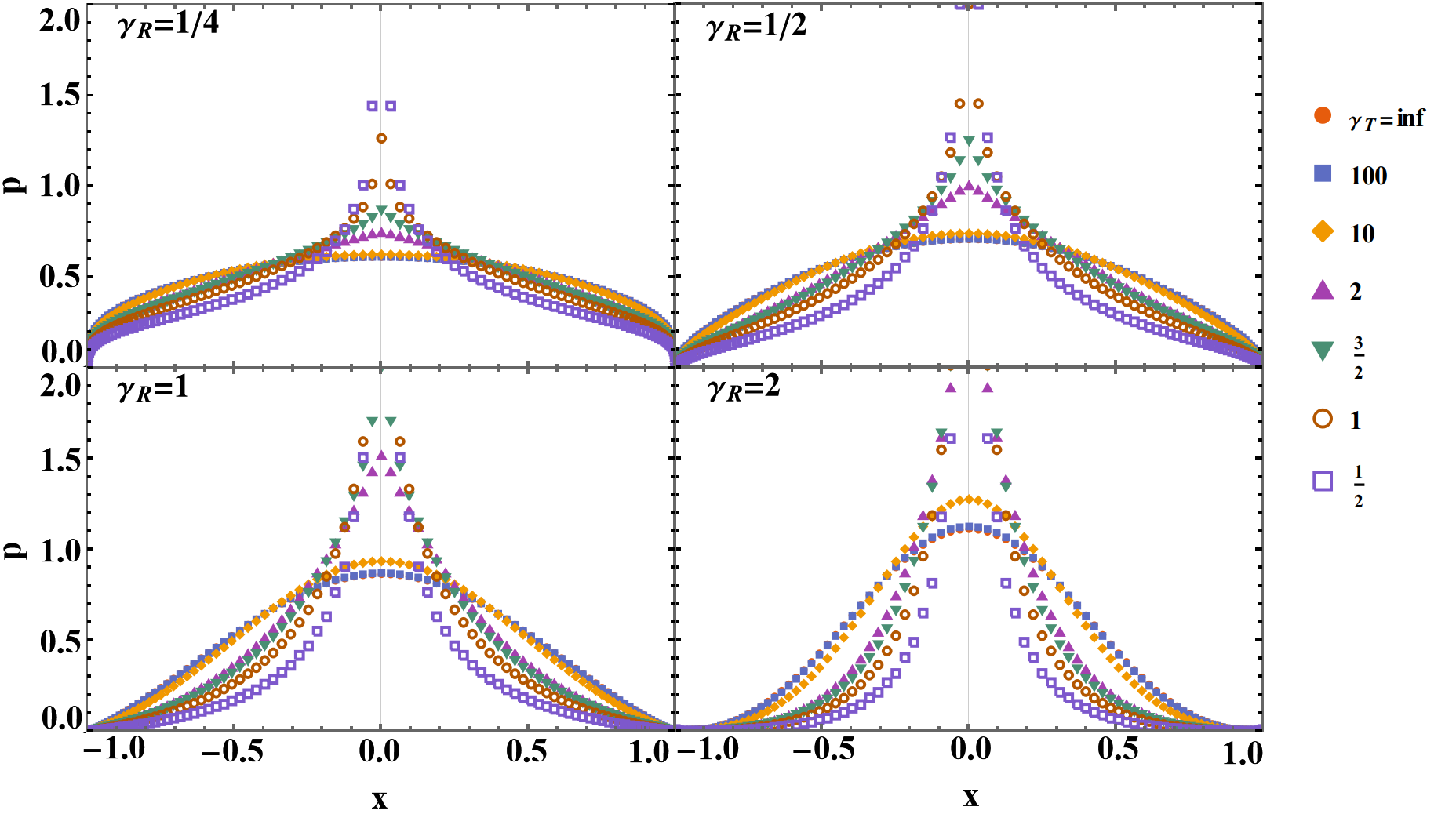} \caption{Approximated distribution from Gaussian quadrature for a RTP in the
harmonic trap with finite tumble time in $3D$ with $N=100$ moments.
The harmonic potential $b$ and the active velocity $v$ are rescaled
to $1$. The run rates are $\gamma_{R}=1/4,1/2,1,2$ respectively.
In each graph the legends gives the value of $\gamma_{T}$.}
\label{fig:3d} 
\end{figure*}

We then proceed to estimate the steady state in higher dimension in
Fig. \ref{fig:2d} and Fig. \ref{fig:3d}. By comparison, the exact
distribution for $\gamma_{T}\rightarrow\infty$ in 2D is also presented
in Fig. \ref{fig:2d}. In 3D, the exact distribution is unknown, although
the moments for $\gamma_{T}\rightarrow\infty$ were also calculated
in \cite{2022_PositingtheProblemofStationaryDistributionsofActiveParticlesAsThirdOrderDifferentialEquation,2023_RunandTumbleOscillatorMomentAnalysisofStationaryDistributions}.
As the result is visually indistinguishable from $\gamma_{T}=100$
here they are not explicitly shown. The behavior near the boundary
agrees with our conclusion in Subsection \ref{subsec:Remarks}. The
behavior near the center is more complicated. Our argument is only
correct if distribution $p$ is smooth near $x_{j}$, thus the points
given at the center may not be correct. Yet still, it seems to suggest
that $\gamma_{T}<1$ will still produce a singular peak.

\section{Conclusion}

\label{sec:Conclusion}

Traditionally the study of many stochastic processes has focused on
the Fokker-Planck equation. The Langevin equation, or the equation
of motion, has been sometimes considered as less suitable for further
development, and consequently few works have actually focused on it.
Yet in this work, it seems fair to claim that at least for some problems,
the equation of motion itself can be quite powerful, leading to results
that are not easily derived from the corresponding Fokker-Planck equation.

Guided by the seminal work of Mark Kac \citep{1974_AStochasticModelRelatedtotheTelegraphersEquation},
we calculated the moments of RTP in a harmonic potential with finite
tumble time. We first formulate the stochastic term and calculate
its correlation function with the law of total expectations. Then
we perform the integral using the Laplace transform to obtain the
moments. These can be summarized in the form of the diagram laws,
and the programmable Volterra difference equation. In 1D, we obtain
the exact steady state distributions, generalizing the previous results
\citep{2020_ExactStationaryStateofaRunandTumbleParticlewithThreeInternalStatesinaHarmonicTrap}
to arbitrary choice of $\gamma_{R}$ and $\gamma_{T}$. In 2D and
3D we extend the results from \citep{2022_PositingtheProblemofStationaryDistributionsofActiveParticlesAsThirdOrderDifferentialEquation,2023_RunandTumbleOscillatorMomentAnalysisofStationaryDistributions}
to the cases with a finite tumble time. To estimate the distributions
from moments, we provide one practical solution to the moment problem
by invoking the Chebyshev polynomials. 

With this method, we are able to determine the behavior introduced
by the finite tumble time in the RTP model, especially in a harmonic
trap. We see that in general, $\gamma_{T}<1$ results in a singular
peak at the center for the steady state distribution. This is to be
expected, as in the tumble state, the particle is pulled towards the
center by the potential, and small tumble rate means longer tumble
time, resulting in more pulling. What is perhaps not expected is that
this behavior is largely independent of the run rate. Similarly, the
tumble time seems to have little effect on the singular peak near
the boundary for small $\gamma_{R}$. The two parameters govern the
behavior near the center and boundary almost independently. 

Nevertheless, questions still remain. The first question is the time
dependent problem. In principle we are solving the problem for any
time, yet at the end, due to the difficulty in obtaining the inverse
Laplace transform, only the infinite time, \textsl{viz.}, steady state
distribution is obtained. It would be interesting to consider other
times as well. The moments are all rational in the Laplace parameter
$\xi$, thus in principle there is no difficulty to obtain the inverse.
In practice, however, the results will be complicated.

Another question is whether such approach can be generalized to some
other problems. One such possibility is that the velocity would be
biased in one direction. Also possible is a more complex Markov chain
for $R$ to model other problems in the general theory of stochastic
phenomena. A good example remains to be found. Furthermore, it would
be interesting to consider a harmonic potential, or to include the
interaction of particles, though it is unclear how to achieve this
since we need the explicit solution to the equation of motion.

Finally, it would be interesting to apply the Gaussian quadrature
method to practical data processing, \textsl{e.g}., inferring the
distribution from given data. As long as we can find a scheme to approximate
any smooth function $f$ supported on the same interval as the distribution
function by grids $x_{j}$ and corresponding cardinal functions $C_{j}$
(where $C_{j}=\sum_{i=0}^{N}J_{ij}T_{i}\left(x\right)$ in our Chebyshev
expansion), our argument seems to hold. Thus by calculating the expectation
of the cardinal functions from the data, it seems possible to estimate
$p\left(x_{j}\right)$. Furthermore, assuming $p$ itself is smooth,
then it can be expanded using the same cardinal functions $C_{j}$,
and the coefficients we need in such expansion are exactly $p\left(x_{j}\right)$
that we just estimated.

\section{Acknowledgments}

The authors would like to thank D. Frydel for his insightful comments
on an earlier version of the manuscript. FY ackowledges the support
of the National Natural Science Foundation of China (Grant No. 12090054
and 12325405) and the Strategic Priority Research Program of Chinese
Academy of Sciences (Grant No. XDB33030300). RP acknowledges the support
of the Key project of the National Natural Science Foundation of China
(NSFC) (Grant No. 12034019).

\appendix

\section{The 2D active Brownian Particles}

While not the focus of this paper, it is worthwhile to mention that
the methodology described here can be adapted to the (2D) \textsl{Active
Brownian particles} (ABP) \cite{2024_ActiveOscillatorRecurrenceRelationApproach}
as well, with active velocities diffusing along a circle. The equation
of motion for the ABP is:

\begin{equation}
\dot{x}\left(t\right)=-bx\left(t\right)+v\cos\theta\left(t\right)
\end{equation}
\begin{equation}
\dot{\theta}\left(t\right)=\sqrt{2D}\eta
\end{equation}
Where $D$ is the diffusion constant and $\eta$ is the standard Gaussian
white noise. We then again use the solution:

\begin{align}
x\left(t\right) & =v\int_{0}^{t}\cos\theta\left(t\right)e^{-b\left(t-s\right)}ds
\end{align}
(with $x\left(0\right)=\theta\left(0\right)=0$) to calculate the
moments: 
\begin{widetext}
\begin{align}
\left\langle r\left(t\right)^{l}\right\rangle = & v^{l}l!\int_{0\leqslant t_{1}\leqslant...\leqslant t_{l}\leqslant t}dt_{i}e^{-b\left(lt-\sum_{k}t_{k}\right)}\left\langle \prod_{k}\cos\theta\left(t_{k}\right)\right\rangle =\frac{v^{l}l!}{2^{l}}\sum_{a_{i}=\pm1}\int_{0\leqslant t_{1}\leqslant...\leqslant t_{l}\leqslant t}dt_{i}e^{-b\left(lt-\sum_{k}t_{k}\right)}\left\langle e^{\sum_{k}ia_{k}\theta\left(t_{k}\right)}\right\rangle \label{eq:momentsabp}
\end{align}
\end{widetext}

Using the standard identity valid for Gaussian variables 
\[
\left\langle e^{A}\right\rangle =e^{\left\langle A\right\rangle +\frac{1}{2}\left(\left\langle A^{2}\right\rangle -\left\langle A\right\rangle ^{2}\right)},
\]
and the correlation function $\left\langle \theta\left(t_{i}\right)\theta\left(t_{j}\right)\right\rangle =2D\min\left(t_{i},t_{j}\right)$,
it can be checked that: 
\begin{equation}
\left\langle e^{\sum_{k}ia_{k}\theta\left(t_{k}\right)}\right\rangle =e^{-D\sum_{k=0}^{l-1}\left(t_{k+1}-t_{k}\right)\left(\sum_{i=k+1}^{l}a_{i}\right)^{2}},
\end{equation}
where we use the convention $t_{0}=0$. Together with the identity
Eq. \ref{eq:integraltoconvolve}, we see that Eq. \ref{eq:momentsabp}
is again a convolution, and the corresponding Laplace transform is:
\begin{equation}
\widetilde{\left\langle x^{l}\right\rangle }\left(\xi\right)=\frac{v^{l}l!}{2^{l}}\sum_{a_{i}=\pm1}\prod_{k=0}^{l}\frac{1}{D\left(\sum_{i=k+1}^{l}a_{i}\right)^{2}+bk+\xi},
\end{equation}
where it is understood that $\sum_{i=l+1}^{l}...=0$. The steady state
moments agree with the examples given in \citep{2024_ActiveOscillatorRecurrenceRelationApproach}.

 \bibliographystyle{unsrt}
\addcontentsline{toc}{section}{\refname}\bibliography{../lib}

\end{document}